\documentclass[aps,pra,twocolumn,superscriptaddress,showpacs]{revtex4}

\usepackage [latin1] {inputenc}
\usepackage {t1enc}
\usepackage [dvips] {graphicx}
\usepackage {verbatim}
\usepackage {epsfig}
\usepackage {amsmath}
\usepackage {amssymb}

\usepackage {psfrag}

\newcommand{\eqnref}[1]{(\ref{#1})}
\newcommand{\figref}[1]{Fig.~\ref{#1}}

\newcommand{\ah}{\hat a}
\newcommand{\sh}{\hat s}
\newcommand{\vacr}{|vac\rangle}

\newcommand{\bra}[1]{\langle #1|}
\newcommand{\ket}[1]{|#1\rangle}

\newcommand{\rubsev}{${}^{87}$Rb}
\newcommand{\rubfiv}{${}^{85}$Rb}
\newcommand{\strsev}{${}^{87}$Sr}

\makeindex

\begin{document}

\title{Fast Entanglement Distribution with Atomic Ensembles and Fluorescent Detection}
%\title{Fast Entanglement Distribution with Atomic Ensembles and Efficient Detection}
\date{\today}

\author{J. B. Brask}\affiliation{QUANTOP, The Niels Bohr Institute, University of Copenhagen, 2100 Copenhagen \O, Denmark}
\author{L. Jiang}\affiliation{Department of Physics, Harvard University, Cambridge, MA 02138, USA}\affiliation{Institute for Quantum Information, California Institute of Technology, Pasadena, CA 91125, USA}
\author{A. V. Gorshkov}\affiliation{Department of Physics, Harvard University, Cambridge, MA 02138, USA}
\author{V. Vuletic}\affiliation{Harvard-MIT Center for Ultracold Atoms, Department of Physics, MIT, Cambridge, MA 02139, USA}
\author{A. S. S\o rensen}\affiliation{QUANTOP, The Niels Bohr Institute, University of Copenhagen, 2100 Copenhagen \O, Denmark}
\author{M. D. Lukin}\affiliation{Department of Physics, Harvard University, Cambridge, MA 02138, USA}

\begin{abstract}
Quantum repeaters based on atomic ensemble quantum memories are promising candidates for achieving scalable distribution of entanglement over long distances. Recently, important experimental progress has been made towards their implementation. However, the entanglement rates and scalability of current approaches are limited by relatively low retrieval and single-photon detector efficiencies. We propose a scheme, which makes use of fluorescent detection of stored excitations to significantly increase the efficiency of connection and hence the rate. Practical performance and possible experimental realizations of the new protocol are discussed.
\end{abstract}


\maketitle

%Introduction------------------------------------------------------

%General quantum comm. and repeater intro 

Distribution of entanglement over long distances has diverse potential applications, ranging from absolutely secure cryptography to fundamental tests of quantum mechanics \cite{gisin}. The task is challenging since direct transmission of quantum states via optical fibres suffers from exponential attenuation, rendering communication beyond a few hundred kilometers impossible. The losses can be overcome by implementing a so-called quantum repeater architechture, which divides a chanel into small segments and combines quantum memories, entanglement swapping and entanglement purification to extend entanglement generated over these segments to longer distances \cite{briegel}. Spurred by the proposal of Ref.~\cite{dlcz} (DLCZ), a number of promising quantum repeater protocols based on storage of light in atomic ensembles have recently been put forward \cite{zhao,jiang,simon,sangouard2}. Apart from atomic ensembles, these schemes require only simple linear optical operations and photodetection. They therefore lend themselves well to experimental realization, and extensive progress has been made towards their implementation \cite{chen2,choi,chaneliere,eisaman,thompson}.

%Problems in prev. schemes, what's new in our scheme

An important limiting factor in all existing atomic ensemble-based quantum repeater schemes is the efficiency of entanglement swapping, which requires conversion from excitations stored in atoms to light followed by single-photon detection. In practice, the combined retrieval and photodetection efficiency is on the order of ten percent \cite{choi,chen2}. This severely limits the communication rates and thus the scalability of the existing approaches. Here we present a new ensemble-based quantum repeater, which circumvents this problem by storing multiple excitations in a single atomic ensemble, as in Ref. \cite{brion}. Entanglement swapping is achieved by fluorescent detection of the populations of certain atomic levels, eliminating the need for retrieval. Fluorescent detection can have very high efficiency and at the same time allows us to determine the number of stored excitations. For trapped ions, fluorescent detection efficiencies of 99.99\% have been experimentally demonstrated \cite{myerson}, and similar techniques have been proposed for photon counting using ensemble-based memories for light \cite{imamoglu,james}. Employing such an idea allows the success probability for entanglement connection to be notably enhanced. Our protocol can be implemented either as a single-rail scheme (analogous to the original DLCZ scheme) or as a dual-rail scheme (analogous to the proposals of Refs.~\cite{jiang,zhao}). Below we will focus on the single-rail scheme for simplicity, but we stress that the protocol may readily be extended to dual-rail, including etanglement purification \cite{brask2}.

%The Scheme------------------------------------------------------

\begin{figure}
\includegraphics[width=.45\textwidth]{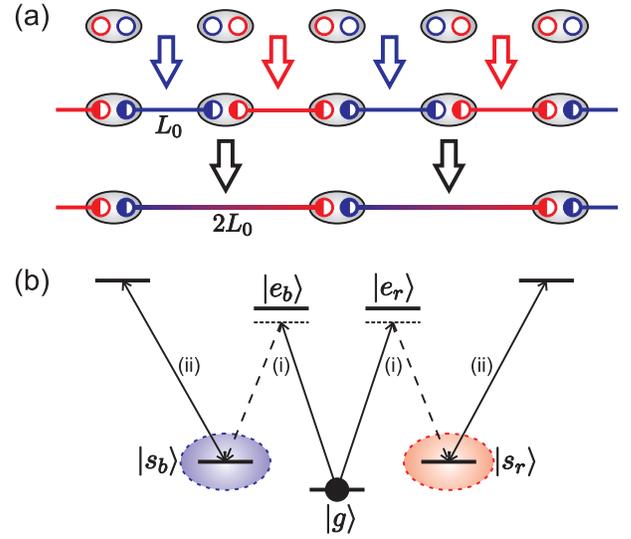}
\caption{(Color online) \textbf{(a)} One atomic ensemble with a 'red' and a 'blue' level makes up each repeater node. At the first step of the protocol, neighboring nodes are entangled with 'red' and 'blue' links established asynchronously. At subsequent steps, distant nodes are connected via entanglement swapping. \textbf{(b)} Atomic level scheme. Encircled levels store spin waves. The transitions (i) and (ii) are employed for entanglement generation and fluorescent detection, respectively.}
\label{fig.schemeoverview}
\end{figure}

\begin{figure}
\includegraphics[width=.45\textwidth]{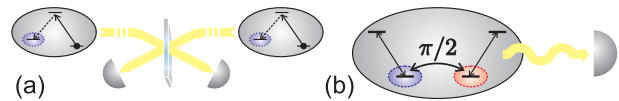}
\caption{(Color online) \textbf{(a)} Entanglement generation with the blue levels. \textbf{(b)} Entanglement connection consisting of a $\pi/2$ pulse between the storage levels followed by fluorescent measurement of their populations.}
\label{fig.entgenandcon}
\end{figure}

As illustrated in \figref{fig.schemeoverview}a, in our approach, repeater nodes seperated by a distance $L_0$ each contain a single ensemble of $N$ atoms encoding two qubits via the level structure shown in \figref{fig.schemeoverview}b. Each atom has a reservoir level, two storage levels, henceforth denoted by `red' and `blue', and at least one cycling transition to an excited state which allows populations to be measured. Connecting to previous ensemble-based repeaters based on $\Lambda$-scheme atoms, one may think of \figref{fig.schemeoverview}b as a double $\Lambda$-scheme -- one for each storage level -- with two additional cycling transitions \cite{dlcz,jiang,zhao}. As discussed below, the proposed level scheme can be implemented in alkali or alkaline earth atoms. Every ensemble is initialised in the `vacuum' state $\ket{g}^{\otimes N}$ with all $N$ atoms in the reservoir. To entangle two ensembles (\figref{fig.entgenandcon}a) we focus first on their blue levels. In each ensemble, a weak laser pulse induces Raman scattering, preparing a joint state of the atoms and the forward scattered Stokes light mode \cite{dlcz}
\begin{equation}
\label{eq.twomodesqueezed}
(1 + \sqrt{q \eta} \, \sh_b^\dagger \ah^\dagger )\vacr + O(\eta q),
\end{equation}
where $\sh_b^\dagger = \frac{1}{\sqrt{N}}\sum_i \ket{s_b}_i\bra{g}$ creates a symmetric atomic spin wave, $\ah^\dagger$ creates a Stokes photon, $\vacr$ is the joint vacuum state of atoms and light, $q$ is the excitation probability and $\eta $ is the fraction of light scattered into the forward mode. The Stokes photons are then mixed on a balanced beam-splitter, and conditioned on the detection of a single click at the output ports, the ensembles are projected to an entangled state of the form % to erase their which-path information
\begin{equation}
\label{eq.maxentstate}
( \sh_{b,1}^\dagger + \sh_{b,2}^\dagger )\vacr ,
\end{equation}
where 1,2 label the ensembles. In this manner entanglement can be established at every other link of the repeater. To entangle the remaining links, the process is repeated using the red levels. For just three ensembles, the resulting state is
\begin{equation}
\label{eq.twoentlinks}
( \sh_{r,2}^\dagger + \sh_{r,3}^\dagger )( \sh_{b,1}^\dagger + \sh_{b,2}^\dagger )\vacr .
\end{equation}
This ideal scenario is implemented if we can avoid multiple excitations of the symmetric spin wave, which is the case when $\eta q \ll 1$. Also note that in addition to forward scattering described by Eq. \eqnref{eq.twomodesqueezed}, excitation of non-symmetric atomic modes occurs with probability $(1-\eta)q$. We analyze the contribution from these excitations below. Once two neighbouring entangled links are established they are connected by entanglement swapping (\figref{fig.entgenandcon}b). A $\pi/2$ rotation is applied between the two storage levels in the central ensemble, and the populations of these levels are then measured by fluorescent detection. Conditioned on the detection of a single excitation in either the red or the blue level, the outermost ensembles are projected to an entangled state. %The $\pi/2$ rotation can be implemented via microwave pulses and radiofrequency magnetic fields, and near-unit efficiencies may be achieved \cite{merkel}. 
Referring to Eq. \eqnref{eq.twoentlinks}, the rotation acts like a beam splitter on the atomic operators taking $\sh_{i,2}^\dagger \rightarrow (\sh_{b,2}^\dagger + (-1)^{\delta_{ir}} \sh_{r,2}^\dagger)/\sqrt{2}$, where $i = b,r$. The subsequent fluorescent detection effectively projects one atom from the central ensemble into $\ket{s_r}$ or $\ket{s_b}$ while the remaining atoms are projected to the reservoir state. As a result, up to a known phase flip ensembles 1 and 3 are projected to the entangled state
\begin{equation}
\label{eq.afterentcon}
( \sh_{b,1}^\dagger + \sh_{r,3}^\dagger )\vacr .
\end{equation}
In the absence of imperfections the probability for a single atom to fluoresce is $2N/(4N-1) \sim 1/2$, and hence the ideal probability for entanglement connection to succeed is $1/2$ as in previous schemes. However, since the DLCZ-protocol requires conversion from atomic to optical excitations and subsequent single-photon detection, it has much higher loss in practice than the present scheme.

%Purification by interrupted retrieval------------------------------------------

Because the same atoms encode several qubits and because of the use of fluorescent detection, several issues not present in previous protocols must be considered in our scheme. First, fluorescent detection does not selectively detect the symmetric spin wave. This is in contrast to schemes based on retrieval, for which collective enhancement ensures that only excitations associated with Stokes photons scattered forward during entanglement generation will be detected during connection \cite{dlcz}. As a consequence, multiexcitation errors occur in our scheme with probability $q$ as opposed to $\eta q$ in previous schemes. To reach a given final fidelity the generation success probability, and thus the rate, must therefore be decreased by a factor of $\eta $, unless corrective measures are taken.

Specifically, to suppress excitations in modes other than the symmetric mode, one can use \textit{purification by interrupted retrieval} (PIR) based on electromagnetically induced transparency. The behavior of a spin wave excitation with momentum $\mathbf{k}$ under retrieval depends on the optical depth $d_\mathbf{k}$ in the direction $\mathbf{k} + \mathbf{k}_c$, where $\mathbf{k}_c$ is the control field wave-vector. The symmetric spin wave has $\mathbf{k}\sim\mathbf{0}$ and for an elongated ensemble, we can arrange that $d_\mathbf{0}=d$, where $d\gg 1$ is the on-axis optical depth. Hence if we were to retrieve the symmetric spin wave from one of the storage levels by applying a control field to the $s \rightarrow e$ transition (dropping subscripts), the retrieved field would travel with the group velocity $v_g = \Omega^2 l/\gamma d$, where $\Omega$ is the Rabi frequency of the control field, $\gamma$ is the decay rate of the excited level, and $l$ is the length of the ensemble \cite{gorshkovII}. In other words, retrieval is highly directional and the field travels at a group velocity inversely proportional to $d$. On the other hand, for excitations with sufficiently small $d_\mathbf{k}$, this picture is not valid. When $d_\mathbf{k}\lesssim 1$, the excitation will decay without any directionality at the rate of a single emitter $\Omega^2/\gamma$, where $\Omega \ll \gamma$. By placing the ensemble in a cavity or inside a hollow-core photonic crystal fibre, as discussed below, we can arrange that only a few modes have high $d_\mathbf{k}$. E.g. in an hollow-core fibre, only the guided and near-transversal modes persist, while intermediate modes are suppressed. It is then possible to turn on the control field for a duration $T$ such that
\begin{equation}
\frac{l}{v_g} \gg T > \frac{\gamma}{\Omega^2}.
\end{equation}
This means that the retrieval is interrupted before the symmetric spin wave leaves the ensemble, while excitations in other modes escape. Clearly there will be a trade-off between loss of the symmetric spin wave and suppression of the incoherent excitations. One can show that the fraction of the symmetric spin wave lost is at most $\delta = 2 v_g T /l$, where half of the loss is simply the retrieved field escaping in the forward direction while the second half comes from spontaneous decay \cite{brask2}. The excitations in all other modes are suppressed by a factor $\eta + (1-\eta)e^{-\delta d/2}$, and thus reducing the multiexcitation error probability in our scheme to $O(\eta q)$ costs $\delta \sim -2\log (\eta ) / d$. The loss probability $\delta$ and the inefficiency of fluorescent detection can be regarded as a connection inefficiency, equivalent to the combined retrieval and detection losses in previous schemes (where retrieval loss scales with $d^{-1/2}$ \cite{gorshkovII}). For reasonable optical depths $d \gtrsim 100$ and a forward scattering probability $\eta$ at or above the percent level, $\delta$ is less than ten percent. This is in contrast to state of the art conventional approaches, in which losses are on the order of 90\% or more. Hence, even with PIR the total connection loss in our scheme is significantly lower than the corresponding losses in schemes with full retrieval.

%Bounds on atom no. ----------------------------------------

Two additional imperfections lead to constraints on $N$. Fluorescent detection implies an upper bound, because measurements will exhibit high dark counts if the number of atoms in the reservoir is too large \cite{imamoglu,james}. Population in the reservoir can contribute to dark counts in two ways. Either through off-resonant scattering on the $g \leftrightarrow e$ transition of light from the probe field, which is resonant with the cycling transition, or through population transfer into $\ket{s}$ caused by the probe. Except when the branching ratio $\beta$ for decay from $\ket{e}$ to $\ket{s}$ is tiny, the latter of these provides a more severe restriction on the atom number, because it is amplified by subsequent resonant scattering. For simplicity we assume that the two transitions have the same dipole moment, such that we can associate a single Rabi frequency $\Omega_p$ with the probe field, and that the decay rates from both excited states are $\gamma$. The fluorescent scattering rate $r$ and the scattering rate $r'$ from $\ket{g}$ into $\ket{s}$ are then given by
\begin{equation}
r = \frac{\gamma\Omega_p^2}{\gamma^2 + 2\Omega_p^2} \hspace{.5cm} \text{and} \hspace{.5cm} r' = \frac{\beta\gamma\Omega_p^2}{4\Delta^2} ,
\end{equation}
where the frequency difference between the $g \leftrightarrow e$ and cycling transitions, denoted by $\Delta$, is much larger than both $\Omega_p$ and $\gamma$. The rates $r$ and $r'$ determine the measurement time and the amount of population transfered from reservoir to storage level during that time, respectively. The time required to faithfully detect a single excitation in the storage level via fluorescent detection is $n/\eta \eta_d r$ where $\eta_d$ is the single photon detection efficiency and $n$ is the desired average number of measured photons. The expected number of logical dark counts, i.e.~the amount of population transferred during the measurement, is
\begin{equation}
\label{eq.darkcountprob}
\frac{Nr'n}{\eta \eta_d r} = \frac{n \beta}{\eta \eta_d} \frac{\gamma^2 + 2\Omega_p^2}{4 \Delta^2} N.
\end{equation}
Clearly, it is desirable to implement our scheme in a system where $\Delta$ can be large, such that high $N$ and thus high $d$ can be reached without introducing dark counts.

A lower bound on $N$ arises because an atom taking part in only the red or the blue spin wave, thus being entangled e.g.~'to the left' but not 'to the right', may lead to an accepted connection with no resulting entanglement. If for a given repeater node one entangled link has been established using e.g.~$\ket{s_b}$, failed entanglement generation attempts on $\ket{s_r}$ degrade the entanglement by projecting atoms to $\ket{s_r}$ or $\ket{g}$, removing them from the blue spin wave. Assuming that undesired population in $\ket{s_r}$ can be removed after a failed attempt (by shelving in a metastable level or heating out of the trap) only decay into $\ket{g}$ will lead to mismatch between the spin waves, when a second entangled link is subsequently established. An average of $1/\beta\eta' q$ attempts are needed to establish the second entangled link, where $\eta' = \eta \eta_d e^{-L_0/2L_{att}}$ and $L_{att}$ is the fibre attenuation length. Each attempt projects $(1-\beta) q$ atoms from the first spin wave into $\ket{g}$ so in total $(1-\beta)/\beta\eta'$ atoms take part only in one spin wave. Taking $\beta\eta' N \gg 1$, the probability to have a separable state after entanglement connection is of order $(1-\beta)/\beta\eta' N$.

%Estimate of the improvement in rate----------------------------------------------------------

\begin{figure}
\includegraphics[width=.45\textwidth]{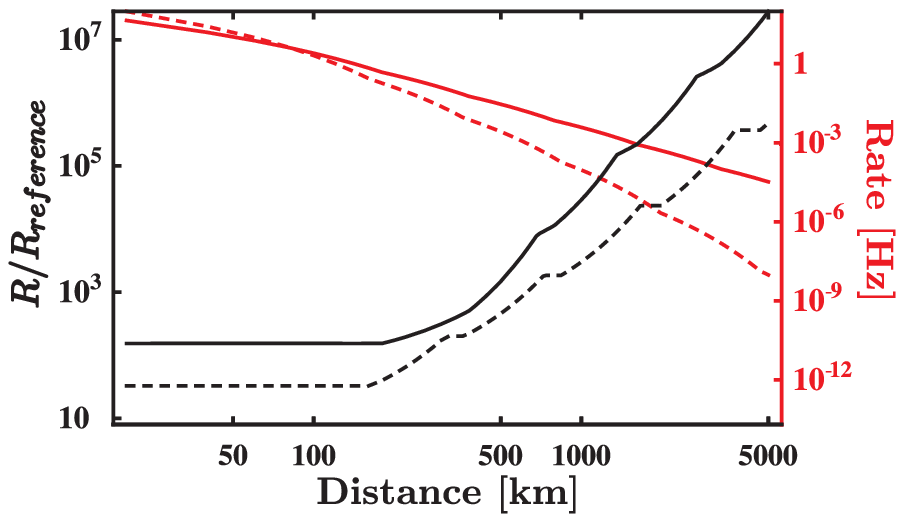}
\caption{(Color online) Ascending curves: Ratio of the rate in the present dual-rail (solid) and single-rail (dashed) scheme with PIR to those of DLCZ and Ref.~\cite{jiang}, assuming $L_{att} = 20$km, $\eta=0.05$, optical depth 100, and efficiencies .95 for fluorescent detection in the present scheme and .4 for single-photon detection. Steps occur when the numbers of nodes change. Descending curves: the corresponding absolute rates in the new scheme. The final fidelity is 90\%.}
\label{fig.rateratio}
\end{figure}

\figref{fig.rateratio} shows an estimate of the improvement in rate when the upper and lower bounds on $N$ are compatible and multiexcitation errors, coming from the higher-order terms in Eq. \eqnref{eq.twomodesqueezed}, is the dominant error. The rates in both the new scheme and the reference schemes have been optimised for the given parameters \cite{suppinfo}. The single-rail estimate is based on the rate expression derived in Ref.~\cite{brask} for the DLCZ scheme, while the dual-rail estimate is based on Ref.~\cite{jiang}. As we have argued above, this is justified for ensembles placed in cavities or hollow-core fibres enabling PIR at the cost of a small increase in connection loss. From the figure we see that significant improvements of two to four orders of magnitude can be expected at distances around 1000km. We note that the attenuation lengths relevant to different quantum repeater schemes may depend on specific implementations. In \figref{fig.rateratio} the attenuation lengths of our scheme and the reference schemes are assumed equal. However, we have verified that the advantage of our scheme persists even when the attenuation lengths differ significantly \cite{suppinfo}.

%Discussion of particular implementations--------------------------------------------

We now discuss specific implementations. We consider an atomic sample confined within a photonic waveguide or a cavity, which enables preferential coupling to a small set of optical modes. Consider e.g. cold alkali atoms such as \rubsev, confined in a single-mode fiber. As demonstrated recently \cite{bajcsy}, such a system can have high optical depth, thus, enabling effective PIR. Specifically, it is feasible, with some improvement in the confinement of both atoms and photons, to reach the regime of $\eta \sim 0.05$ and $d \sim 100$ with $2\cdot 10^3$ atoms. Fluorescence measurements can be done on the D2-line cycling transition, which has $\gamma = 6$MHz, and the reservoir can be separated from the detected level by the ground state hyperfine splitting $\Delta = 6.8$GHz. Taking, for instance, $\eta=0.05$, $\eta_d=0.5$, $n=20$, and $\beta=0.5$, we find that with a single connection, an entangled state between nodes separated by 10km can be created. Taking atoms with a higher nuclear spin, e.g.~\rubfiv, such entangled ensembles can be used as a backbone for implementing the entire repeater chain, provided that advanced \cite{zhao,jiang,simon,sangouard2} protocols including purification are employed.

Higher fidelity implementations can be achieved with alkaline earth atoms. Alkaline earths are very well suited for implementation due to the presence of long-lived metastable levels which make it possible to separate the cycling transition from any transitions involving the reservoir by optical frequencies. E.g.~in \strsev, by cycling on the ${}^1S_0 \leftrightarrow {}^1P_1$ transition while keeping the reservoir temporarily in ${}^3P_0$ we can have $\Delta \approx 10$THz, while $\gamma = 30$MHz \cite{brask2}. In addition it can be arranged that measurements do not induce population transfer from the reservoir into the cycling transition, such that only off-resonant scattering can contribute to dark counts. This means that the factor $n\beta/\eta \eta_d$ can be dropped from Eq. \eqnref{eq.darkcountprob}, relaxing the upper bound further. At the same time, if excitation and decay ${}^3P_0 \rightarrow {}^1P_1 \rightarrow {}^1S_0$ \cite{reichenbach} is used for entanglement generation, the mismatch error can be suppressed by a very large branching ratio $1-\beta \sim 10^{-8}$, so the lower bound is simply $1/\eta'$. For a repeater over 1000km with $2^5$ segments and final fidelity $F > 95\%$, we estimate an upper bound of $\sim 10^8$ while the lower bound above gives $\sim 1$ \cite{brask2}. Thus, a repeater might be implemented with $N \sim 10^4$ atoms compatible with good optical depth. We note that while the upper bound for alkali earths is high enough to allow large $d$ to be reached in free space, PIR requires that only a few nearly symmetric modes see high optical depth. Thus even for large $N$ it is desireable to enclose the ensemble in a low-finesse cavity, enhancing $d$ for the symmetric mode, or in a single-mode hollow-core fibre, since without PIR our rate would be suppressed by a factor $\eta$, which is small in free space. Another promising candidate for implementation may be cavity enclosed crystals of ions with long-lived excited levels, e.g.~Ca${}^{+}$ or Sr${}^{+}$. Strong collective coupling in such a system has recently been demonstrated \cite{herskind}.

%Conclusion------------------------------

In summary, we have described a new ensemble-based quantum repeater protocol that uses fluorescent detection to significantly improve the efficiency of entanglement swapping. Our scheme can be implemented with atoms in a single-mode hollow-core fibre or a low-finesse cavity and yields improvements in communication rate for a distance of 1000km by two to four orders of magnitude over previous proposals. An interesting extension of our scheme would be to incorporate non-linearities such as Rydberg blockade \cite{brion}.

%Acknowledgements

We acknowledge helpful discussions with I. Cirac, E. Polzik, M. Bajcsy and S. Hofferberth. This work was supported by NSF, CUA, DARPA, the EU FET-Open project COMPAS (212008), and the Danish National Research Foundation.

\end{document}